\documentclass[a4paper,11pt]{article}
\pdfoutput=1
\usepackage[utf8x]{inputenc}
\usepackage[left=2.8cm,right=2.8cm,top=2.8cm,bottom=2.8cm]{geometry}
\usepackage[linktoc=all,hidelinks]{hyperref}
\usepackage{cite}
\usepackage{amsmath}
\usepackage{amsfonts}
\usepackage{amssymb}
\usepackage{slashed}

\newcommand{\pd}{\partial}

\newcommand{\nn}{\nonumber}

\newcommand{\cC}{\mathcal{C}}

\newcommand{\cM}{\mathcal{M}}

\newcommand{\cO}{\mathcal{O}}
\newcommand{\cV}{\mathcal{V}}
\newcommand{\cX}{\mathcal{X}}

\newcommand{\id}{\mathbf{1}}
\DeclareMathOperator{\ch}{ch}
\DeclareMathOperator{\tr}{tr}
\DeclareMathOperator{\Tr}{Tr}

\DeclareMathOperator{\Sym}{Sym}

\newcommand{\AM}{\widehat{A}(\cM)}

\begin{document}
\numberwithin{equation}{section}

\thispagestyle{empty}
\begin{center}

\vspace*{30pt}
{\LARGE \bf
Gravitational anomalies of \\\vspace{10pt}
fermionic higher-spin fields}

\vspace{30pt}
{Victor Lekeu${}^{\,a}$ and Yi Zhang${}^{\, b, \,c}$}

\vspace{10pt}
\texttt{victor.lekeu@aei.mpg.de, yi.zhang@ipht.fr}

\vspace{20pt}
\begin{enumerate}
\item[${}^a$] {\sl \small
Max-Planck-Institut für Gravitationsphysik (Albert-Einstein-Institut)\\
Am Mühlenberg 1, 14476 Potsdam, Germany}

\item[${}^b$] {\sl \small
Center for High Energy Physics, Peking University, Beijing 100871, China}

\item[${}^c$] {\sl \small
Institut de Physique Théorique, Université Paris Saclay, CNRS, CEA\\
91191 Gif-sur-Yvette Cedex, France}

\end{enumerate}

\vspace{30pt}

\begin{abstract}
Using the Atiyah-Singer index theorem, we formally compute gravitational anomalies for fermionic higher-spin fields in two, six and ten dimensions, as well as the $U(1)$ mixed gauge-gravitational anomaly in four dimensions. In all cases, anomaly cancellations are found for an infinite tower of fields with alternating chiralities.
\end{abstract}

\end{center}

\vspace{25pt}




\section{Introduction}

We consider gravitational anomalies for chiral fermionic symmetric higher-spin fields. These are massless tensor-spinor fields $\psi^\alpha_{\mu_1 \dots \mu_s}$ of rank $s$ which are symmetric in their space-time indices and describe, in four space-time dimensions, higher-spin particles of half-integer helicity $h =  s + 1/2$ \cite{Fang:1978wz}. The free theory can only be formulated on a flat or AdS background, and these fields are notoriously difficult to couple to dynamical gravity; c.f.~\cite{Aragone:1979hx} for the spin $5/2$ case and \cite{Bekaert:2010hw} for a useful review of no-go theorems. However, there is by now growing evidence that higher-spin gravity theories should in fact exist and are of a non-conventional nature, containing an infinite tower of massless fields of all spins. We refer to \cite{Bekaert:2022poo} for a recent review of the field.

In any putative higher-spin theory containing chiral fermions coupled to gravity, an important consistency requirement at the quantum level is the vanishing of the total gravitational anomaly. We compute here the local gravitational anomalies of such fields on an arbitrary Euclidean manifold $\cM$ in two, six and ten dimensions, as well as the $U(1)$ mixed gauge-gravitational anomaly in four dimensions. This is done using the Atiyah-Singer index theorem \cite{Atiyah:1970ws,Atiyah:1971rm}, following \cite{Alvarez-Gaume:1983ihn,Alvarez-Gaume:1983ict,Alvarez-Gaume:1984zlq,Alvarez:1984yi}. We treat this method as a `formal black box' which takes as only input the field and ghost content of the theory and returns the anomaly polynomial, a $(D+2)$-form written in terms of the Pontryagin classes of $\cM$ (with $D = \dim{\cM}$). We do not address the problem of whether such a theory can be constructed on the manifold $\cM$ or not. Instead, it is hoped that anomaly cancellations based on the spectrum of fields can give necessary conditions for their existence. This computation therefore comes with some assumptions:
\begin{enumerate}
\item We assume that the number of independent gauge transformations of the field is the same as in flat or AdS space, where the free action has been constructed long ago \cite{Fang:1978wz}.
\item We assume that the link between the anomaly and the index density computed with the theorem still holds; in particular, it is unclear to us how non-minimal couplings of higher-spin fields to gravity affect this result, if at all (for conventional fields, they do not \cite{Alvarez-Gaume:1983ihn}).
\end{enumerate}
The first of these assumptions is very reasonable. Further investigation of the second assumption, in particular using the explicit couplings of chiral higher-spin gravity \cite{Ponomarev:2016lrm,Skvortsov:2018jea}, is left for future work. It would also be interesting to repeat this calculation with the slightly different field and ghost content of \emph{conformal} higher spins, and check the result using their couplings to Weyl gravity \cite{Fradkin:1985am,Fradkin:1990ps,Segal:2002gd} (see also \cite{Tseytlin:2013jya}).

Despite these caveats, we find a very suggestive result, namely: in every case, the (regularised) total anomaly for an infinite tower of fields with alternating chiralities vanishes.

\section{The index theorem}

The anomaly polynomial $\hat{I}^{(s)}_{D+2}$ is given by the $(D+2)$-form part of the index density of a certain Dirac operator \cite{Alvarez-Gaume:1983ihn,Alvarez-Gaume:1983ict,Alvarez-Gaume:1984zlq,Alvarez:1984yi} which is written in the form
\begin{equation} \label{eq:Diracstandard}
\slashed{D} : \mathcal{C}^{\infty} (S^+ \otimes \cV_s) \longrightarrow \mathcal{C}^{\infty}(S^- \otimes \cV_s)\, ,
\end{equation}
where $S^+$ (resp. $S^-$) denotes the positive (resp. negative) spinor representation of $SO(D)$ and $\cV_s$ is a formal sum of tensor representation spaces of $SO(D)$ carrying the information about the index structure of the field and ghosts. Then, using the index theorem \cite{Atiyah:1970ws,Atiyah:1971rm} one has
\begin{equation}\label{eq:index}
\hat{I}^{(s)}_{D+2} = \left[\text{Ind}(\slashed{D})\right]_{D+2} = \left[ \ch(-\mathrm{F}) \AM  \ch(R,\cV_s) \right]_{D+2}\, .
\end{equation}
The subscript indicates that one has to pick out the $(D+2)$-form part of this expression, which is a sum of forms of even degrees. We now briefly explain the various components of this formula, and refer to \cite{Bilal:2008qx,Lekeu:2021oti} for a pedagogical introduction.

\paragraph{Chern character of the gauge group.} First of all, if the field transforms in a representation $\mathfrak{v}$ of a Yang-Mills gauge group, one has the Chern character
\begin{equation}
    \ch(-\mathrm{F}) = \tr_\mathfrak{v} \exp \left( -\frac{i\mathrm{F}}{2\pi}\right) = \dim(\mathfrak{v}) - \frac{i}{2\pi} \tr_\mathfrak{v}(\mathrm{F}) - \frac{1}{2(2\pi)^2} \tr_\mathfrak{v}(\mathrm{F}^2) + \dots
\end{equation}
where traces are taken in the representation $\mathfrak{v}$, and $\mathrm{F} = F^a T^{(\mathfrak{v})}_{a}$ is the curvature two-form contracted with the generators of the gauge group in that representation. The factor $\ch(-\mathrm{F})$ is thus a sum of forms of even degree.

\paragraph{Dirac genus.} The second factor is the Dirac genus (or roof genus) $\AM$ of the manifold: it is a sum of forms of degrees that are multiple of four and reads, up to the twelve-form component,
\begin{align}
\AM &= 1- \frac{1}{24} \,p_1 + \frac{1}{5760}(7 \, p^2_1 - 4 \, p_2) + \frac{1}{967680} \left( -31 \, p_1^3 + 44 \, p_1 p_2 - 16 \, p_3 \right) + \dots\, .
\end{align}
This is written in terms of the Pontryagin classes of the manifold $\cM$; each $p_k$ is a form of degree $4k$. Up to $p_3$, they are given in terms of the Riemann curvature two-form $R^a{}_b$ by
\begin{align}
p_1 &= \frac{1}{(2 \pi)^2} \left(-\frac{1}{2} \tr R^2 \right) \label{eq:p1}\\
p_2 &=\frac{1}{(2 \pi)^4} \left(-\frac{1}{4} \tr R^4 +\frac{1}{8} (\tr R^2)^2 \right) \label{eq:p2} \\
p_3 &= \frac{1}{(2 \pi)^6} \left(-\frac{1}{6} \tr R^6 +\frac{1}{8} \tr R^2 \tr R^4 - \frac{1}{48} (\tr R^2)^3 \right) \, . \label{eq:p3}
\end{align}
In flat or (A)dS space, these classes (and hence the gravitational anomaly) identically vanish. That is not the case for a generic manifold $\cM$, for which we will nevertheless find interesting anomaly cancellations in the next section.

\paragraph{Chern characters of $SO(D)$ and trace identities.} To evaluate the last factor $\ch(R,\cV_s)$ in formula \eqref{eq:index}, we will need the Chern character of the rank $s$ symmetric tensor representation of $SO(D)$. It is defined as
\begin{align}
\ch(R_{(s)}) &= \Tr \exp\left( \frac{i R_{(s)}}{2\pi} \right) \nn\\
&= \frac{(D+s-1)!}{s!(D-1)!} - \frac{1}{2!} \frac{1}{(2\pi)^2} \Tr (R_{(s)}^2) +  \frac{1}{4!} \frac{1}{(2\pi)^4} \Tr(R_{(s)}^4) + \dots \, . \label{eq:chernsymmetric}
\end{align}
Here, $R_{(s)}$ is given in terms of the curvature two-form of the manifold by $R_{(s)} = \frac{1}{2} R_{ab} T_{(s)}^{ab}$, where the $T_{(s)}^{ab}$ are the generators of the rank $s$ symmetric representation of $SO(D)$:
\begin{align}\label{eq:Ts}
(T_{(s)}^{ab})^{i_1 \dots i_s}{}_{j_1 \dots j_s} = s! \sum_{l=1}^s \delta^{(i_1}_{j_1} \delta^{i_2}_{j_2} \cdots (t^{ab})^{i_l}{}_{j_l} \cdots \delta^{i_s)}_{j_s}\, ,
\end{align}
with $(t^{ab})^i{}_j = 2 \delta^{i[a} \delta^{b]}_j$ the generators of the fundamental (vector) representation. In formula \eqref{eq:chernsymmetric}, we used $\Tr(\id) = \binom{D+s-1}{s} = \frac{(D+s-1)!}{s!(D-1)!}$ and the fact that traces of odd powers of $R_{(s)}$ identically vanish here. Thus $\ch(R_{(s)})$, like $\AM$, is a sum of forms of degrees that are multiple of four. For our purposes (up to $D=10$), we need explicit formulas up to the twelve-form component, i.e.~$\Tr(R_{(s)}^6)$. 

As a next step, one needs to write the traces of powers of $R_{(s)}$ in terms of Pontryagin classes or, equivalently, in terms of traces in the fundamental representation. This is achieved by expanding the generating function (see e.g.~\cite{Okubo:1983sv,Schellekens:1986xh})
\begin{align}\label{eq:generatingformula}
\sum_{s=0}^\infty x^s \ch(R_{(s)}) = \det\left(1 - x \, e^{\frac{i R}{2\pi}}\right)^{-1} = \exp\left[ - \tr \log \left(1 - x \, e^{\frac{i R}{2\pi}}\right) \right]\, .
\end{align}
We find the following explicit formulas, for arbitrary rank $s$ and dimension $D$. For the four-form, one has simply
\begin{align} \label{eq:QD2}
    \Tr (R_{(s)}^2) &= Q_2(s) \tr R^2\, , \quad Q_2(s) = \frac{(D+s)!}{(s-1)!(D+1)!} \, .
\end{align}
For the eight-form, two types of terms are possible: $\tr(R^4)$ and $( \tr R^2 )^2$. Indeed, one finds
\begin{align} \label{eq:QD4}
    \Tr (R_{(s)}^4) &= Q_4(s) \tr R^4 + 3\, A_4(s) ( \tr R^2 )^2
\end{align}
where $Q_4(s)$ and $A_4(s)$ are
\begin{align}
Q_4(s) &= \frac{(D+s)!}{(D+3)! (s-1)!} \left(6 s^2 + (6s-1) D + D^2 \right) \\
    A_4(s) &= \frac{(D+s+1)!}{(D+3)! (s-2)!}\, .
\end{align}
Finally, for the twelve-form one has the following combination of the three possible terms $\tr R^6$, $\tr R^2 \tr R^4$ and $( \tr R^2 )^3$:
\begin{align}
    \Tr (R_{(s)}^6) &= Q_6(s) \tr R^6 + 15\, A_6(s) \tr R^2 \tr R^4 + 15\, B_6(s) ( \tr R^2 )^3
\end{align}
with
\begin{align}
    Q_6(s) &= \frac{(D+s)!}{(D+5)! (s-1)!} \Big(120 s^4 + \left(240 s^3-90 s^2+4\right) D \nn \\ 
    &\qquad\qquad\qquad\qquad\qquad + \left(150 s^2-90 s+11\right) D^2 + (30 s-16) D^3 + D^4\Big) \\
    A_6(s) &= \frac{(D+s+1)!}{(D+5)! (s-2)!} \left(6 s^2-4 + (6 s-3) D + D^2 \right)\\
    B_6(s) &= \frac{(D+s+2)!}{(D+5)! (s-3)!} \, .
\end{align}
These formulas can equivalently be written in terms of Pontryagin classes, using \eqref{eq:p1} -- \eqref{eq:p3}.

\paragraph{Field and ghost content.} Finally, the object $\cV_s$ is a formal sum of $SO(D)$ representation spaces, determined by the field and, after quantisation, ghost content of the theory. The result is
\begin{equation}\label{eq:Vs}
    \cV_s = \Sym^{s} T^*\!\cM - \Sym^{s-1} T^*\!\cM - \Sym^{s-2} T^*\!\cM + \Sym^{s-3} T^*\!\cM\, ,
\end{equation}
so the factor $\ch(R,\cV_s)$ appearing in \eqref{eq:index} is defined as
\begin{equation}
    \ch(R,\cV_s) = \ch(R_{(s)}) - \ch(R_{(s-1)}) - \ch(R_{(s-2)}) + \ch(R_{(s-3)})\, .
\end{equation}
For $s \leq 2$, one should omit the terms with a negative index.

We now indicate how to derive this result. For definiteness, we take the field $\psi_{\mu_1 \dots \mu_s}$ to be of positive chirality (for a negative-chirality field, the anomaly simply gets an extra minus sign). First of all, in the Fang-Fronsdal ``metric-like'' formulation \cite{Fang:1978wz}, the field $\psi^\alpha_{\mu_1 \dots \mu_s}$ and the gauge parameter $\zeta^\alpha_{\mu_1 \dots \mu_{s-1}}$ satisfy the gamma-trace conditions
\begin{align}
    \gamma^\mu\gamma^\nu\gamma^\rho \psi_{\mu\nu\rho\sigma_4 \dots \sigma_s} = \eta^{\mu\nu}\gamma^\rho \psi_{\mu\nu\rho\sigma_4 \dots \sigma_s} = 0\, , \quad \gamma^\mu \zeta_{\mu\nu_3 \dots \nu_s} = 0\, .
\end{align}
Therefore, the field $\psi_{\mu_1\dots \mu_s}$ can be seen as an element of the formal difference
\begin{equation}
    \cC^\infty(S^+ \otimes \Sym^{s} T^*\!\cM - S^- \otimes \Sym^{s-3} T^*\!\cM)\, ,
\end{equation}
that is, a positive-chirality, symmetric tensor-spinor of rank $s$, without (since the chirality matrix anticommutes with gamma matrices) its \emph{negative}-chirality rank $s-3$ component. To unclutter notation, we will write
\begin{equation}
    (s)_\pm = \cC^\infty(S^\pm \otimes \Sym^{s} T^*\!\cM)
\end{equation}
from now on for the space of rank $s$ symmetric tensor-spinor fields of positive or negative chirality, without gamma-trace condition. So, $\psi \in (s)_+ - (s-3)_-$ and, similarly, $\zeta \in (s-1)_+ - (s-2)_-$.

We will impose the gauge condition $\chi(\psi) = 0$, where
\begin{equation} \label{eq:gaugecondition}
    \chi_{\mu_2 \dots \mu_s}(\psi) \equiv \gamma^{\mu_1}\psi_{\mu_1 \mu_2 \dots \mu_s}- \frac{s-1}{D+2(s-2)} \eta^{\nu\rho}\gamma_{(\mu_2} \psi_{\mu_3 \dots \mu_s)\nu\rho}\, .
\end{equation}
This condition satisfies $\gamma^{\mu_2} \chi_{\mu_2 \dots \mu_s}(\psi) = 0$ identically, which is consistent with the gamma-tracelessness of the gauge parameter $\zeta$. To understand what this does, let us denote by $\hat{\psi}^{(s)}$ a gamma-traceless, symmetric tensor-spinor of rank $s$. The triple gamma-trace condition on the field $\psi$ means that it only contains components of type $\hat{\psi}^{(s)}$, $\hat{\psi}^{(s-1)}$ and $\hat{\psi}^{(s-2)}$; imposing the gauge condition $\chi(\psi) = 0$ then sets $\hat{\psi}^{(s-1)} = 0$, leaving $\hat{\psi}^{(s)}$ and $\hat{\psi}^{(s-2)}$.

Now, in the usual delta-function gauge-fixing (after integrating out auxiliary fields), one only integrates over fields $\psi$ satisfying the gauge condition $\chi(\psi) = 0$. There is also a pair $(\hat{C}^{(s-1)},\hat{\bar{C}}^{(s-1)})$ of Faddeev-Popov ghosts: they are positive-chirality, symmetric, gamma-traceless tensor spinors of rank $s-1$ just as the gauge-parameter $\zeta$, but they are bosonic fields (with wrong spin-statistics). Together, the fields and ghosts are then an element of the formal space
\begin{align}
    \cX_\delta &= \left[ (s)_+ - (s-1)_- \right] + \left[ (s-2)_+ - (s-3)_- \right] - 2 \left[ (s-1)_+ - (s-2)_- \right]\, ,
\end{align}
where the first term corresponds to $\hat{\psi}^{(s)}$, the second to $\hat{\psi}^{(s-2)}$, and the third to the two ghosts $(\hat{C}^{(s-1)},\hat{\bar{C}}^{(s-1)})$ which come with a negative sign because of the wrong spin-statistics. Now, since a negative-chirality spinor contributes to the index density with an extra sign compared to a positive-chirality spinor, we can effectively write $\cX_\delta$ as
\begin{align}
    \cX_\delta &= (s)_+ + (s-1)_+ + (s-2)_+ + (s-3)_+ - 2 (s-1)_+ - 2 (s-2)_+  \nn\\
    &= (s)_+ - (s-1)_+ - (s-2)_+ + (s-3)_+\, .
\end{align}
Using these formal rules, $\cX_\delta$ has therefore been brought in the canonical form $\cC^\infty(S^+ \otimes \cV_s)$ required in \eqref{eq:Diracstandard}, where $\cV_s$ is indeed as in \eqref{eq:Vs}.

As a cross-check, one could also quantise the system with Gaussian gauge-fixing. The gauge-breaking term then takes the schematic form $\chi \cdot \slashed{\pd} \chi$, with an extra  differential operator to ensure that the quadratic part of the action is homogeneous in derivatives as is customary for spin $3/2$. Then, a third ghost $\hat{b}^{(s-1)}$ is needed, the Nielsen-Kallosh ghost \cite{Nielsen:1978mp,Kallosh:1978de} (see also \cite{Batalin:1983ar,Lekeu:2021oti} for a general description in the Batalin-Vilkovisky quantisation formalism). It satisfies the same properties as the gauge condition $\chi(\psi)$: it is a symmetric, gamma-traceless tensor spinor of rank $s-1$, of the opposite chirality as $\psi$. Importantly, it is also a fermionic field with the correct spin-statistics. In this scheme, the two Faddeev-Popov ghosts have opposite chiralities but otherwise identical properties and their contribution to the index density cancels. The fields and ghosts are then an element of
\begin{align}
    \cX_\text{G} &= \left[ (s)_+ - (s-3)_- \right] + \left[ (s-1)_- - (s-2)_+ \right] \nn \\
    &= (s)_+ - (s-1)_+ - (s-2)_+ + (s-3)_+\, ,
\end{align}
i.e., $\cX_\text{G}=\cC^\infty(S^+ \otimes \cV_s)$ with the same $\cV_s$, as expected: the choice of quantisation scheme should not affect the result.

\section{Gravitational anomalies}

With these formulas in hand, we now compute the pure gravitational anomalies in two, six and ten spacetime dimensions, and the $U(1)$ mixed gauge-gravitational anomaly in four dimensions. In all cases, the total anomaly polynomial is found to cancel when summed over all values of $s$ with alternating signs.

\paragraph{The two-dimensional case.} 

In $D=2$ and up to the four-form component, the Chern character $\ch(R_{(s)})$ reduces to
\begin{equation}\label{eq:2dchern}
    \ch(R_{(s)}) = (s+1) + \frac{1}{6} s(s+1)(s+2) p_1 + \dots\, .
\end{equation}
The anomaly polynomial is then found to be
\begin{align}
\hat{I}^{(s)}_{4} &= \left[ \AM  \ch(R,\cV_s) \right]_{4} \nn\\
&= \left[ \left( 1- \frac{1}{24} p_1 \right) \Big( \ch(R_{(s)}) - \ch(R_{(s-1)}) - \ch(R_{(s-2)}) + \ch(R_{(s-3)})\Big)\right]_{4} \nn\\
&= (2s-1)\, p_1 \, .\label{eq:anomaly2d}
\end{align}
This formula is valid for all $s \geq 2$. For $s=0$ and $1$, the complex $\cV_s$ is smaller and one gets
\begin{align}
    \hat{I}^{(0)}_{4}  = \left[ 1- \frac{1}{24} p_1 \right]_{4}  = - \frac{1}{24} p_1\, , \quad
    \hat{I}^{(1)}_{4} =  \left[\left(1- \frac{1}{24} p_1 \right) \left( 1 + p_1 \right) \right]_{4} = \frac{23}{24} p_1
\end{align}
as expected \cite{Alvarez-Gaume:1983ihn}. For $s=2$, formula \eqref{eq:anomaly2d} is still valid since formally $\ch(R_{(-1)}) = 0$ in \eqref{eq:2dchern}.

With this result in hand, we now investigate anomaly cancellations for a tower of higher-spin fields. If we consider one field for each integer $s$, all fields having the same chirality, we are led to the infinite sum $\sum_{s=2}^\infty (2s - 1)$ over odd numbers, which diverges and needs to be regularised. There are several ways to do this. One rather elegant way is to analytically continue the identity
\begin{equation}
    \sum^{\infty}_{s=1}(2s-1)^{-\tau} = (1-2^{-\tau}) \zeta(\tau)\, ,
\end{equation}
valid for $\text{Re}(\tau) > 1$. The right-hand side has a smooth limit when $\tau$ goes to $-1$:
\begin{equation}\label{eq:2dreg1/12}
    (1-2^{-\tau}) \zeta(\tau) \xrightarrow{\tau \to {-1}} (1-2) \zeta(-1) = \frac{1}{12}\, ,
\end{equation}
thus assigning the value $1/12$ to the divergent sum. Another way to regularise the sum is to introduce an exponential regulator $e^{-\epsilon s}$, and to drop the poles: this gives
\begin{equation}\label{eq:2dreg1/3}
    \sum^{\infty}_{s=1}(2s-1) e^{-\epsilon s} = \frac{2}{\epsilon^2}-\frac{1}{\epsilon}+\frac{1}{3}+\cO(\epsilon)\, ,
\end{equation}
assigning this time the value $\frac{1}{3}$ to the divergent sum. (This is of course equivalent to the naive zeta function regularisation $\sum^{\infty}_{s=1}(2s-1) \to 2 \zeta(1) - \zeta(0) = \frac{1}{3}$.) Slightly more generally, one can use the regulator $e^{-\epsilon (s + x)}$, as was done in \cite{Giombi:2014iua,Gunaydin:2016amv} with simple dimension-dependent values for the extra parameter $x$, giving
\begin{equation}\label{eq:2dsamechirality}
    \sum^{\infty}_{s=1}(2s-1) e^{-\epsilon (s + x)} = \text{(singular)} + \frac{1}{3} \left(1 + 3x + 3 x^2 \right)+\cO(\epsilon)\, .
\end{equation}
This agrees with \eqref{eq:2dreg1/12} for $x = -1/2$, with \eqref{eq:2dreg1/3} for $x=0$, and can also produce many other values. Clearly, there is a big ambiguity and some extra physical input is required to choose a natural regularisation. Let us note, however, that the choice $1/12$ leads to
\begin{align}
    \sum^{\infty}_{s=0} \hat{I}_4^{(s)} &= \left( -\frac{1}{24} + \frac{23}{24} + \sum^{\infty}_{s=2}(2s-1) \right) p_1 \nn\\
    &\rightarrow  \left( -\frac{1}{24} + \frac{23}{24} + \frac{1}{12} -1 \right) p_1 \nn\\
    &= 0 \, .\label{eq:2dtotalanomalies}
\end{align}
Remarkably, the situation is much less ambiguous when one considers an alternating sum. Then, the exponential regulator $e^{-\epsilon s}$ gives an expression that is regular as $\epsilon \to 0$:
\begin{equation}
    \sum^{\infty}_{s=2} (-1)^s (2s - 1) e^{-\epsilon s} = 1 + \cO(\epsilon)\, .
\end{equation}
(One does not need to ``drop the poles''; the finite part is unambiguous.) Moreover, there is no extra parameter in this scheme: using $e^{-\epsilon (s+x)}$ clearly does not change the result. Now, since the anomaly of a right-handed spinor is the opposite of that of a left-handed one, this formula then leads to a cancellation of anomalies for a tower of fields with alternating chiralities:
\begin{align}
    \sum_{s=0}^{\infty} (-1)^s \hat{I}^{(s)}_{4} e^{-\epsilon s} &= \left( -\frac{1}{24} - \frac{23}{24} e^{-\epsilon} + \sum^{\infty}_{s=2} (-1)^s (2s - 1) e^{-\epsilon s} \right) p_1 \nn\\
    &= \left( -\frac{1}{24} - \frac{23}{24} + 1 \right) p_1 + \cO(\epsilon) \nn\\
    &= 0 + \cO(\epsilon)\, .
\end{align}

\paragraph{Mixed gauge-gravitational anomaly in four dimensions.}

There is no purely gravitational anomaly in four spacetime dimensions. However, there are mixed gauge-gravitational anomalies: using \eqref{eq:index}, the complete anomaly polynomial is given by the six-form
\begin{align}
\hat{I}^{(s)}_{6} &= \left[ \ch(-\mathrm{F}) \AM \ch(R,\cV_s) \right]_{6}\nn\\
&= - \frac{i}{2\pi} \left(\tr \mathrm{F}\right) \left[ \AM \ch(R,\cV_s) \right]_{4} + \frac{i}{6(2\pi)^3} \tr(\mathrm{F}^3)\, .
\end{align}
The second term, proportional to $\tr(\mathrm{F}^3)$, is the usual gauge anomaly for chiral fermions: it is insensitive to the index structure of the field and takes the same value here as in the spin $1/2$ case. The first term is the mixed gauge-gravitational anomalies. Only the $U(1)$ factors can contribute and we have $\tr \mathrm{F} = - i q_s F$, where $F = dA$ is the $U(1)$ gauge connection and $q_s$ the charge of the field (if there are several $U(1)$ factors, one gets a sum of such terms). The second factor is evaluated as before: one needs the Chern character $\ch(R_{(s)})$ up to the four-form component,
\begin{equation}
    \ch(R_{(s)}) = \frac{1}{6}(s+1)(s+2)(s+3) + \frac{1}{120}s(s+1)(s+2)(s+3)(s+4) p_1 + \dots\, .
\end{equation}
Putting things together, the  mixed $U(1)$-gravitational anomaly is
\begin{align}
\hat{I}^{(s)\mathrm{M}}_{6} &= -\frac{q_s}{2\pi} F \;\left[ \left( 1- \frac{1}{24} p_1 \right) \Big( \ch(R_{(s)}) - \ch(R_{(s-1)}) - \ch(R_{(s-2)}) + \ch(R_{(s-3)})\Big)\right]_{4}\, \nonumber \\
&= - \frac{1}{24} (1+2s) (-1 +4s +4s^2) \frac{q_s}{2\pi} F p_1 \, .
\end{align}
This formula is valid for all $s \geq 0$ (including $s=0$, $1$ and $2$). This result, being in four dimensions, is of course not entirely new \cite{Christensen:1978md,Romer:1979bh,Marcus:1985yy}; however, anomaly cancellation for an infinite tower of fields was not investigated in these early works.

We now investigate anomaly cancellations for a tower of higher-spin fields.
As a first step, let us consider a tower of fields with equal charges, $q_s = Q_1$ for all $s$, and with the same chirality. Using the exponential regulator $e^{-\epsilon(s+x)}$ yields 
\begin{equation}\label{eq:4dsamechirality}
    \sum_{s=0}^{\infty} \hat{I}^{(s)\mathrm{M}}_{6} e^{-\epsilon (s+x)} =  \text{(singular)} - \frac{1}{5!}\left(10 x^4-20 x^3+5 x^2+5 x-3\right) \frac{Q_1}{2\pi} F p_1 + \cO(\epsilon)\, .
\end{equation}
This can be made to vanish after dropping the singular part, but this requires picking a rather unnatural and \emph{ad hoc} value for the parameter $x$ (which is not of the simple forms considered in \cite{Giombi:2014iua}). 

As in the two-dimensional case, however, the alternating sum is much better behaved. Then, one finds simply
\begin{equation}
    \sum_{s=0}^{\infty} (-1)^s \hat{I}^{(s)\mathrm{M}}_{6} e^{-\epsilon (s+x)} = 0 + \cO(\epsilon)\, .
\end{equation}
This is regular as $\epsilon \to 0$, and independent of $x$. The alternating sum can be arranged by alternating the chirality of the field, as before, or by keeping the same chirality but alternating the sign of the charge. Furthermore, we find that having all charges equal is not the only possibility: for example, taking
\begin{equation}
q_s = Q_1 + Q_2\, s (s+1)
\end{equation}
(still with alternating signs) also gives zero in this regularisation scheme. It would be interesting to know the most general function $q_s$ that still leads to a cancellation.

\paragraph{The six-dimensional case.} 

For the anomaly polynomial in six spacetime dimensions, we find the eight-form
\begin{align}
    \hat{I}^{(s)}_{8} &= \left[ \AM  \ch(R,\cV_s) \right]_{8} \nn\\
    &= \frac{1}{5760 \times 42} (s+1) (s+2) (2 s+3) \left[ C_1(s) \, p^2_1 - C_2(s) \, p_2\right]
\end{align}
where $C_1(s)$ and $C_2(s)$ are the fourth-degree polynomials
\begin{align}
C_1(s) &= 48 s^4+288 s^3+324 s^2-324 s+49 \\
C_2(s) &= 48 s^4+288 s^3+576 s^2+432 s+28 \,.
\end{align}
This formula is valid for all $s \geq 0$, and reproduces the classic results of \cite{Alvarez-Gaume:1983ihn} for $s=0$ and $1$.

Summing this expression over all values of $s$ using the regulator $e^{-\epsilon(s+x)}$ as before, one has the same behaviour as in \eqref{eq:4dsamechirality}: the finite part is a polynomial in $x$, which can be made to vanish only by picking ``by hand'' an unnatural value of the parameter $x$. However, with alternating chiralities the regularised sum vanishes:
\begin{equation}
    \sum_{s=0}^{\infty} (-1)^s \hat{I}^{(s)}_{8} e^{-\epsilon s} = 0 + \cO(\epsilon)\, .
\end{equation}

\paragraph{The ten-dimensional case.} 

In ten dimensions, we find the following formula for the anomaly polynomial, valid for all $s \geq 0$: 
\begin{align}
    \hat{I}^{(s)}_{12} =& \;\left[ \AM  \ch(R,\cV_s) \right]_{12} \nn\\
    =& \;\frac{1}{967680 \times 720720} \,(s+1) (s+2) (s+3) (s+4) (s+5) (s+6) (2 s+7)\nn \\
    &\times \left[ E_1(s) \, p^3_1 - 2 E_2(s) \, p_1 p_2 + E_3(s) \, p_3\right] \, ,
\end{align}
with
\begin{align}
E_1(s) &= 112 s^6+2352 s^5+15736 s^4+28224 s^3-36015 s^2-2401 s-4433 \,, \\
E_2(s) &= 112 s^6+ 2352 s^5+ 16828 s^4+ 43512 s^3+ 9758 s^2-56546  s-3146 \,, \\
E_3(s) &= 112 s^6+ 2352 s^5+ 17920 s^4+ 58800 s^3+ 68544 s^2- 19600 s- 2288\, .
\end{align}
The sum over all values of $s$ displays the same features as in the lower-dimensional cases: with all fields of the same chirality, the vanishing of the sum requires an unnatural choice of the regulator parameter $x$; with alternating chiralities, one gets the remarkable cancellation
\begin{equation}
    \sum_{s=0}^{\infty} (-1)^s \hat{I}^{(s)}_{12} e^{-\epsilon s} = 0 + \cO(\epsilon)\, .
\end{equation}

\section{Conclusion}

In this short note, we have computed the anomaly polynomials for fermionic higher-spin fields in various dimensions using the Atiyah-Singer index theorem. This requires the computation of some interesting $SO(D)$ trace identities, and careful consideration of the ghost content of the theory. There are of course some assumptions in this computation, detailed above, but nevertheless we find a remarkable result: the total anomaly polynomial vanishes when one considers an infinite tower of fields of all spins with alternate chiralities. The sum over all spins was performed with the natural regulator $e^{-\epsilon s}$, which gives a smooth result as $\epsilon \to 0$.

This work is only a starting point for the investigation of higher-spin gravitational anomalies and should be extended. First of all, it is unclear if non-miminal couplings to gravity affect the result; this should be confirmed using the explicit complete couplings available in \cite{Ponomarev:2016lrm}, and also in the conformal higher-spin case \cite{Fradkin:1985am,Fradkin:1990ps,Segal:2002gd}. Another question is whether higher-spin multiplets containing a tower of fermions with alternating chiralities exist and can be constructed; in fact, in dimensions $D \geq 6$ it would be interesting to repeat the computations of this paper in the case of fermions of mixed symmetry and see whether similar cancellations can be arranged. Finally, in higher dimensions there are chiral \emph{bosonic} higher-spin fields which would contribute to the anomaly; these are rectangular mixed symmetry fields that generalise the self-dual $p$-forms familiar from supergravity \cite{Bekaert:2009fg}. Gravitational anomalies of such fields are yet to be computed (see however \cite{Minasian:2020vxn} for the two-column case).

\subsection*{Acknowledgements}

We would like to thank Evgeny Skvortsov, Xavier Bekaert, Arkady Tseytlin and the anonymous referee for useful discussions, their comments on the first draft/version of this paper, and for pointing out some of the relevant literature to us. YZ would like to thank the Institut de Physique Théorique (IPhT), CEA/Saclay for hospitality. The work of VL is supported by the European Research Council (ERC) under the European Union’s Horizon 2020 research and innovation programme, grant agreement No. 740209. The work of YZ is supported by National Science Foundation of China under Grant No. 12175004, by Peking University under startup Grant No. 7100603534 and by the International Postdoctoral Exchange Program of the Office of China Postdoc Council (OCPC) and Peking University under  Grant No. YJ20220018.


\providecommand{\href}[2]{#2}\begingroup\raggedright\endgroup

\end{document}